\documentclass[a4paper,epj,nopacs]{svjour}
\usepackage[reqno]{amsmath}
\usepackage{amsfonts,amssymb,latexsym}
\usepackage{graphicx}

\newcommand{\One}{1\kern-4.5pt1}

\newcommand{\lapprox}{\raisebox{-0.5ex}{$\
\stackrel{\textstyle<}{\textstyle\sim}\ $}}
\newcommand{\gapprox}{\raisebox{-0.5ex}{$\
\stackrel{\textstyle>}{\textstyle\sim}\ $}}

\newlength{\colw}
\begin{document}
\bibliographystyle{h-physrev4}

\title{Lattice Study of Dense Matter with Two Colors and Four Flavors}

\author{Simon Hands\inst{1}  \and Philip Kenny\inst{1}
\and Seyong Kim\inst{2} \and Jon-Ivar
  Skullerud\inst{3}}

\institute{Department of Physics, College of Science, Swansea University,
 Singleton Park, Swansea SA2 8PP, UK
\and Department of Physics, Sejong University, Gunja-Dong, Gwangjin-Gu,
Seoul 143-747, South Korea
\and Department of Mathematical Physics, National University of Ireland
Maynooth, Maynooth, County Kildare, Ireland}

\abstract{
We present results from a simulation of SU(2) lattice gauge theory with $N_f=4$
flavors of Wilson
fermion and non-zero quark chemical potential $\mu$, using the same $12^3\times24$
lattice,
bare gauge coupling, and pion mass in cut-off units as 
a previous study \cite{Hands:2010gd} with $N_f=2$.  
The string tension for $N_f=4$ is found to be considerably smaller implying
smoother gauge field configurations.
Thermodynamic observables and order parameters for superfluidity and color
deconfinement are studied, and comparisons drawn between the two theories.
Results for quark density and pressure as functions of $\mu$
are qualitatively similar for $N_f=2$ and
$N_f=4$; in both cases there is evidence for a phase in which baryonic matter is
simultaneously degenerate and confined. Results for the stress-energy tensor,
however, suggest that while $N_f=2$ has a regime where dilute matter is
non-relativistic and weakly-interacting, $N_f=4$ matter is relativistic
and strongly-interacting for all values of $\mu$ above onset.
} \maketitle

\section{Introduction}

A first principles QCD-based understanding of the behaviour of dense strongly
interacting matter would be of tremendous value, in part because of the
theoretical interest in new phases of matter exhibiting properties such as color
superconductivity and crystallisation, and also because a quantitatively
accurate equation of state ε$\varepsilon(n_q)$, $p(n_q)$ (where 
$\varepsilon$  is energy density, $p$ pressure
and $n_q$ quark density) is a necessary input for the relativistic equations
of stellar structure determining the properties of compact stellar objects.
Lattice simulations, in principle the most reliable systematic non-perturbative
approach to QCD, are unfortunately unavailable for dense matter, since in the
presence of a quark chemical potential $\mu\not=0$  biasing the system to have more
quarks than anti-quarks, the Euclidean action is complex-valued making Monte
Carlo importance sampling inoperable. Our approach to the problem is to study
QC$_2$D, ie. Yang-Mills plus fundamental quarks with gauge group SU(2), 
which can be shown to have a
real positive functional integral measure for an even number of quark flavors
$N_f$. It is possible using entirely orthodox lattice QCD simulation methods to
generate representative gluon field configurations and study thermodynamic
properties of this model for arbitrary $\mu$ - QC$_2$D is thus the simplest theory in
which dense matter with long-ranged inter-quark interactions can be
systematically studied. 

Previous work has shown that as $\mu$ is raised baryonic matter is induced to
form in the ground state at an onset chemical potential $\mu_o={1\over2}m_\pi$,
since the lightest baryon in the physical spectrum of QC$_2$D is degenerate with the pion
~\cite{Hands:2000ei,Kogut:2001na,Aloisio:2000rb}.
At the same point the U(1)$_B$ baryon-number symmetry of the model is
spontaneously broken by a superfluid diquark
condensate~\cite{Kogut:2001na,Aloisio:2000rb,Hands:2001ee}. Studies of the hadron
spectrum confirm that in this regime there is a Goldstone excitation formed from
a superposition of $qq$ and $\bar q \bar q$ bound states
~\cite{Hands:1999md,Kogut:2001na,Hands:2007uc}.
There is also evidence for a change in the level ordering of $\pi$- and
$\rho$-meson states as density increases~\cite{Muroya:2002ry,Hands:2007uc}.
Subsequent simulations
probing higher densities have found evidence for degenerate quark matter and
color deconfinement~\cite{Hands:2006ve,Hands:2010gd}. There have also been
interesting studies of the glueball spectrum~\cite{Lombardo:2008vc} 
and the presence of gauge field
fluctuations with non-trivial topology~\cite{Alles:2006ea,Hands:2011hd} in a dense medium.

It is, however, difficult to assess this body of work systematically, because
the results have been obtained with differing lattice fermion formulations
(staggered, Wilson), differing numbers of quark flavors ($N_f=2,4,8$), and even
differing matter representations (fundamental, adjoint). Moreover, 
away
from the continuum limit (lattice spacing $a\to0$) staggered lattice fermions 
have global invariances and symmetry breaking patterns 
distinct from those of the continuum theory; 
this has a profound effect in theories
such as QC$_2$D having
real or pseudoreal matter representations, even influencing the nature of the
Sign Problem (see eg. \cite{Hands:2000ei}).

In this work we attempt to remedy the situation by presenting results for
$N_f=4$ obtained with the same lattice action used in previous studies with
$N_f=2$~\cite{Hands:2006ve,Hands:2010gd}. It is intrinsically interesting to 
examine the influence of varying the matter content on the theory's ground
state, especially since $N_f/N_c>1$ in this case, although
it should be immediately noted that the first two coefficients of the
renormalisation group beta-function remain negative, implying the model remains
asymptotically-free and confining for $\mu,T=0$~\footnote{Note that $N_f=4$ lies
below the perturbative 
prediction $N_f^{BZ}\geq{272\over49}$ for the existence of a
conformal fixed point.}.
We will see as $\mu$ is varied that some features of the $N_f=4$ theory are
qualitatively similar to those found for $N_f=2$, while others, particularly the
behaviour of the stress-energy tensor $T_{\mu\nu}$, appear qualitatively very
different. Another important difference is that for the same bare lattice
gauge coupling $\beta$ and pion mass $m_\pi a$ (ie. measured in cut-off units), 
the string tension $\sigma a^2$ is much smaller for $N_f=4$. Strictly, since
the two lattice models yield distinct continuum theories, this cannot
be interpreted in terms of a finer lattice; nonetheless the resulting gauge
field configurations are markedly smoother for $N_f=4$ than for $N_f=2$,
implying that the influence of at least some lattice artifacts will be
diminished. In a follow-up paper \cite{Hands:2011hd} we will demonstrate that this makes the
identification of topological excitations much easier to control, allowing their
behaviour in a dense medium to be probed.

In Sec.~\ref{sec:setup}
we will detail the lattice action used, and give
results for the string tension measured at $\mu=0$. The main results for quark
density $n_q(\mu)$, pressure $p(\mu)$, energy density $\varepsilon(\mu)$ and
trace of the stress energy tensor $T_{\mu\mu}(\mu)$ are presented in
Sec.~\ref{sec:numres}, as well as details of both superfluid and Polyakov line
order parameters yielding information on the physical nature of the matter which
forms. We will find that the main claim of \cite{Hands:2010gd}, namely that
there is a range of $\mu$ in which baryonic matter in QC$_2$D is at once degenerate (ie.
having a well-defined Fermi-surface) and confined, is supported by the results
of the current study. Both similarities and differences between $N_f=4$ and
$N_f=2$ are discussed in Sec.~\ref{sec:disc}.

\section{Simulation Details}
\label{sec:setup}

The lattice action used, as in previous studies in this
series~\cite{Hands:2006ve,Hands:2010gd}, employs Wilson fermions
interacting
with gauge fields governed by the Wilson gauge action.
In units where the lattice spacing $a=1$, 
\begin{equation}
S= \sum_{i,j=1}^2\bar\Psi_x^i{\cal M}_{xy}^{ij}[U;\mu]\Psi_y^j
-{\beta\over
N_c}\sum_{x,\nu<\lambda}\mbox{tr}U_{\nu\lambda x},
\label{eq:S}
\end{equation}
where $U_{\nu\lambda}$ is the oriented product of 4
SU(2)-valued link fields $U_{\nu x}$ around the sides of an elementary plaquette
in the $\nu$-$\lambda$ plane,
and $\mu$ is the quark
chemical potential. The fields $\Psi$ and $\bar\Psi$ are $8N_c$-component
spinors located on the lattice sites; if we write $\Psi^i=(\varphi,\phi)^i$, 
then
\begin{equation}
{\cal M}=\delta^{ij}\left(
\begin{matrix} M[U;\mu] & \kappa j\gamma_5\\
-\kappa j\gamma_5& M[U;-\mu]
\end{matrix}\right)
\end{equation}
with $M[U;\mu]$ the standard Wilson fermion matrix given by
\begin{eqnarray}
M_{xy}[U;\mu]=\delta_{xy}
&-&\kappa\sum_\nu\biggl[(1-\gamma_\nu)e^{\mu\delta_{\nu
0}}U_{\nu x}\delta_{y,x+\hat\nu}\nonumber\\
&+&(1+\gamma_\nu)e^{-\mu\delta_{\nu 0}}U_{\nu
y}^\dagger\delta_{y,x-\hat\nu}\biggr].
\label{eq:M}
\end{eqnarray}
With the identifications $\psi_1=\varphi$, $\bar\psi_1=\bar\varphi$,
$\psi_2=(\bar\phi C\tau_2)^{tr}$,
$\bar\psi_2=(C\tau_2\phi)^{tr}$, where $C\gamma_\mu C^{-1}=-\gamma_\mu^*$ and
$\tau_2$ acts on color, the action (\ref{eq:S}) is readily seen to
be equivalent to two copies of an action describing a quark isodoublet $(\psi_1,\psi_2)$
with the usual coupling to gauge fields, with in addition a gauge invariant 
scalar isoscalar diquark
coupling (or Majorana mass term) of the form
\begin{equation}
\kappa
j(-\bar\psi_1C\gamma_5\tau_2\bar\psi_2^{tr}+\psi_2^{tr}C\gamma_5\tau_2\psi_1).
\label{eq:diquark}
\end{equation}
A diquark source $j\not=0$ mitigates long-wavelength fluctuations and
hence critical slowing down in any
superfluid phase characterised by
$\langle\bar\varphi\gamma_5\phi\rangle\not=0$. It explicitly breaks the global U(1)$_B$
symmetry $\varphi\mapsto e^{i\alpha}\varphi$, $\phi\mapsto e^{-i\alpha}\phi$ 
of (\ref{eq:S})\footnote{In the limit $j=\mu=0$ the full global symmetry group
of the action (\ref{eq:S}) is Sp(8).}. Alternative choices of diquark operator
consistent with the Pauli Exclusion Principle are possible, but
in such cases numerical simulations show no firm evidence for symmetry
breaking leading to superfluidity~\cite{Kenny:2010}. In any case, 
the ``physical'' limit $j\to0$ is
potentially as technically and computationally challenging as the chiral limit in vacuum QCD.

It is possible to show that $\mbox{det}{\cal M}$ is real and positive and hence
that the model can be simulated using an orthodox hybrid Monte Carlo
algorithm~\cite{Hands:2006ve}. In this first study with $N_f=4$, our strategy is
to compare with results obtained for $N_f=2$, using the same $12^3\times24$ lattice and
bare gauge coupling $\beta=1.9$~\cite{Hands:2010gd}. Since in principle
the two theories have distinct physical properties,
there is some arbitrariness in identifying an appropriate matching condition. In 
QC$_2$D the onset transition at which quark density first rises from zero
signalling the transition from vacuum to baryonic matter at $T=0$ is expected
at $\mu_o={1\over2}m_\pi$~\cite{Kogut:2000ek}. Accordingly we have
chosen to match the pion mass measured {\em in lattice units\/}, ensuring that
the onset happens at a similar value of $\mu a$ in the two simulations.
We varied $\kappa$ whilst keeping fixed $\beta=1.9$, $\mu=j=0$, measured
the pion propagator using a local source and sink, and found that $\kappa=0.158$
yielded $m_\pi a=0.677(14)$ in good agreement with the value
$m_\pi a=0.68(1)$ used in the $N_f=2$ study~\cite{Hands:2010gd}.

\begin{figure}[htb]
\includegraphics*[width=\colw]{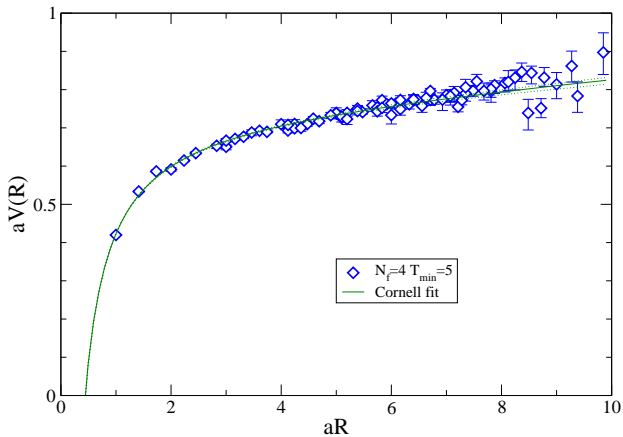}
\caption{The static quark potential from fits to Wilson loops with $aT\geq5$,
together with a fit to the Cornall potential (\ref{eq:Cornell}).
}
\label{fig:4flpot}
\end{figure}
With no further freedom, in order to set the physical scale
we next measured Wilson loops with $R\times T$ with all possible spatial
separations $R$. 
Fitting $W(R,T)$ to the
form $Ae^{-V(R)T}$ for $aT_{\min}\geq5$ (see Fig.~\ref{fig:4flpot}), we use the Cornell potential
\begin{equation}
V(R)=C+{b\over R}+\sigma R
\label{eq:Cornell}
\end{equation}
to extract the string tension $\sigma$. We find $\surd\sigma a=0.1096(64)$,
yielding a lattice spacing $a=0.052(3)$fm assuming a physical string tension
(440 MeV)$^2$\!\!, about a factor of three smaller than the value $a=0.186(8)$fm
found for $N_f=2$~\cite{Hands:2010gd}. Some insight may be gained by integration
of the renormalisation group 
beta-function $\beta(g)=-a{{\partial g/\partial a}}$:
\begin{equation}
a=C\left({g^2\over{16\pi^2b_0+b_1g^2}}\right)^{-{b_1\over{2b_0^2}}}
\exp\left(-{{8\pi^2}\over{b_0g^2}}\right);
\label{eq:2loop}
\end{equation}
with the two-loop coefficients given by
\begin{equation}
b_0={11\over3}N_c-{2\over3}N_f;\;\;
b_1={34\over3}N_c^2-{10\over3}N_cN_f-\left({{N_c^2-1}\over N_c}\right)N_f.
\end{equation}
\begin{figure}[htb]
\includegraphics*[width=\colw]{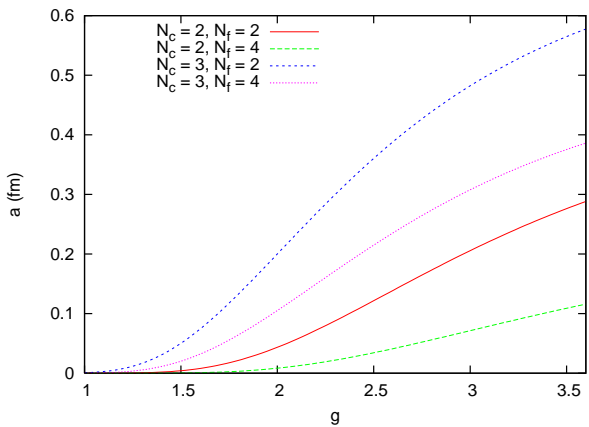}
\caption{\em Plot of the RG equation solution
(\ref{eq:2loop}) for $N_c=2,3$ and $N_f=2,4$. }
\label{fig:dabydg}
\end{figure}
Fig.~\ref{fig:dabydg} plots the solution (\ref{eq:2loop}) for $N_c=2,3$ and
$N_f=2,4$. Whilst it is difficult to directly match the bare lattice coupling 
$\sqrt{2N_c/\beta}$ to
the Yang-Mills coupling $g$ plotted along the horizontal axis, it can be seen
that with two colors at the ``boosted'' value 
$g\approx2.9$, the effect of doubling the number of flavors has the same
dramatic effect of reducing $a$ by a factor of three; in fact the constant $C$
in (\ref{eq:2loop}) can be chosen so that the vertical scale of
Fig.~\ref{fig:dabydg} matches our estimates 
for $a$ made using the string tension $\sigma$. For $N_c=3$ with the same
coupling, by contrast,
changing $N_f$ from 2 to 4 reduces $a$ by just 35\%. The small lattice
spacing at $N_f=4$ is therefore not so surprising, and underlines that changing
the flavor content in QC$_2$D is not as innocent as might naively be thought.
It
also challenges our identification of a physical scale using the string tension
(particularly since we don't have the Particle Data Group to help in QC$_2$D).
Taking $\surd\sigma=440$MeV yields $m_\pi\simeq2.6$GeV and the temperature of
the lattice $T\simeq160$MeV, which on the face of it makes comparison with the
results of \cite{Hands:2010gd} difficult. 

Without data at different values of
$a$ enabling a continuum extrapolation, the best we can achieve in this inital
study of flavor-rich dense matter is therefore a comparison of $N_f=2,4$ in
pion-mass units, since
the two simulations have matched values of
$L_sm_\pi,\,L_tm_\pi$, and in the $T\to0,\,j\to0$ limits should
manifest the onset of baryonic matter
at the same value of $\mu_o/m_\pi$ 
(recall also that from both theoretical~\cite{Kogut:2000ek}
and simulational~\cite{Hands:2000ei} standpoints,
the best way to present
data obtained for $\mu\gapprox\mu_o$ 
with different quark masses uses units of $\mu/m_\pi$). 
As a result, the figures plotted in
Sec.~\ref{sec:numres} all start to exhibit non-trivial behaviour at the
same point along the $\mu$-axis.

We should bear in mind that at $N_f=4$ the gauge field configurations
are likely to be significantly smoother and hence lattice artifacts smaller. 
A potential worry is that the $N_f=4$ simulations are much more
susceptible to finite-volume artifacts and thermal effects as a result of the
much smaller values of $L_s\surd\sigma$ and $L_t\surd\sigma$. 

\section{Numerical Results at $\mu\not=0$}
\label{sec:numres}

The action (\ref{eq:S}) was used to generate between 300 - 500 HMC trajectories
of mean length 0.5 for chemical potentials $\mu a\in[0.25,1.2]$, with diquark
source $ja=0.04$ throughout. The molecular dynamics timestep needed to maintain
75\% acceptance ranged from 
$dt=0.004$ at $\mu a=0.25$ to $dt=0.002$ at $\mu a=1.0$.
\begin{figure}[htb]
\includegraphics*[width=\colw]{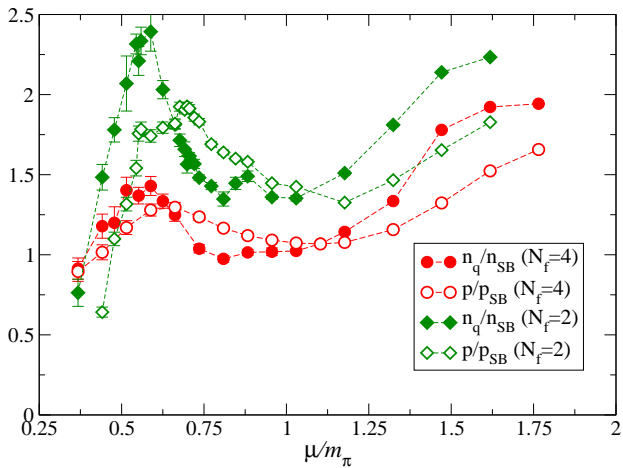}
\caption{\em  $n_q/n_{SB}^{\rm latt}$ and $p/p_{SB}^{\rm latt}$ vs.
$\mu/m_\pi$ for $N_f=2,4$.}
\label{fig:4fln}
\end{figure}
Fig.~\ref{fig:4fln} shows results for the primary thermodynamic observable,
the quark density $n_q(\mu)\equiv-\partial\ln{\cal Z}/\partial\mu$, 
for both $N_f=4$ and $N_f=2$~\cite{Hands:2010gd}.
What is actually plotted is the dimensionless ratio $n_q/n_{SB}^{\rm latt}$,
where $n_{SB}^{\rm latt}$ is the quark density evaluated on the same
$12^3\times24$ lattice with free massless quarks, ie. with
$\beta=\infty,\kappa={1\over8}$. The purpose of this is 
to compensate for both finite-volume effects and lattice artifacts~\cite{Hands:2006ve}.
Note that for $T=0$ in the thermodynamic and continuum limits,
\begin{equation}
n_{SB}^{\rm cont}(\mu)={{N_fN_c}\over{3\pi^2}}\mu^3,
\end{equation}
characteristic of a degenerate system with $E_F=k_F=\mu$.

The most striking feature of Fig.~\ref{fig:4fln} is the qualitative similarity
of the $N_f=2$ and $N_f=4$ data when expressed in cut-off units. In both cases
$n_q/n_{SB}^{\rm latt}$ rises sharply from near the theoretical onset at $\mu_o\approx
{1\over2}m_\pi=0.34a^{-1}$ to a maximum at $\mu/m_\pi \approx0.6$ before falling.
This non-monotonic behaviour is predicted in chiral perturbation theory
($\chi$PT) \cite{Kogut:2000ek,Hands:2006ve}, which treats the matter forming at
onset as a Bose-Einstein condensate of deeply bound scalar isoscalar diquark
pairs.
In fact beyond the maximum, $\chi$PT
predicts the ratio to
be monotonically falling, proportional to $\mu^{-2}$ as $\mu\to\infty$. By
contrast,
at a value $\mu=\mu_Q\approx0.75m_\pi$ the measured ratio levels off; moreover, for $N_f=4$ 
the ratio is close to unity over the range $0.75\lapprox\mu/m_\pi \lapprox1.0$.
Finally, for $\mu>\mu_D\approx1.0m_\pi$ the ratio starts to rise again.

In \cite{Hands:2010gd} the behaviour for $\mu\in(\mu_Q,\mu_D)$
was interpreted as a regime where the
fermionic nature of the quarks has become manifest, resulting in the formation
of a Fermi surface only weakly perturbed by diquark condensation (see below).
In other words, the threshold value $\mu_Q$ corresponds to a BEC/BCS
crossover. The new data at $N_f=4$ strengthens this interpretation, since
the ratio is now so close to one; indeed $n_q/n_{SB}^{\rm latt}$ seems to
approach unity from above as $a\to0$. Since, however, care must be
exercised when extrapolating to the 
continuum limit using data from distinct theories, and 
also because the necessary $j\to0$ extrapolation is still to be done, 
we cannot say at this stage whether any physical significance should be 
attached to the 
differing heights of the peaks at $\mu/m_\pi=0.6$. The flatness of
the $N_f=4$ data in this regime does however increase our confidence in the rather {\it
ad hoc\/} procedure for mitigating lattice artifacts by taking the ratio of
interacting to free theories, which can only strictly be justified if a
continuum limit is taken. 

Fig.~\ref{fig:4fln} also plots the pressure $p$ for both theories, obtained by
integrating the Maxwell relation $n_q=\partial p/\partial\mu$.
For the lattice data we have 
implemented this via the formula~\cite{Hands:2006ve}
\begin{equation}
{p\over{p_{SB}^{\rm latt}}}(\mu)=
{{\int_{\mu_o}^\mu{{n_{SB}^{\rm cont}}\over{n_{SB}^{\rm
latt}}}(\mu^\prime)n_q(\mu^\prime)d\mu^\prime}\over
{\int_{\mu_o}^\mu n_{SB}^{\rm cont}(\mu^\prime)d\mu^\prime}},
\end{equation}
with the integrals estimated by an extended trapezoidal rule.
Since the data for $p$  derive from those for $n_q$,
most of the qualitative comments made above apply here also. We
note that the $N_f=4$ curve is smoother, and that once again the ratio
$p/p_{SB}$ is close to unity in the range $\mu\in(\mu_Q,\mu_D)$.
 
\begin{figure}[htb]
\includegraphics*[width=\colw]{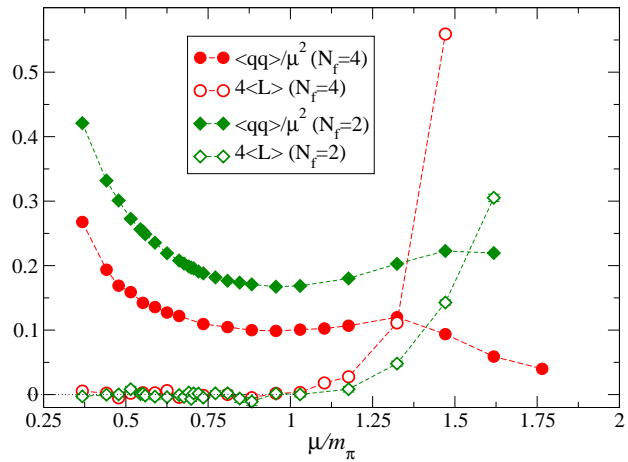}
\caption{\em  $\langle qq\rangle/\mu^2$ and rescaled Polyakov line
vs. $\mu/m_\pi$ for $N_f=2,4$.}
\label{fig:polyandqq}
\end{figure}
In order to characterise the properties of the different regions better, in
Fig.~\ref{fig:polyandqq} we plot two further important observables,
the superfluid order parameter $\langle qq\rangle\equiv
{\kappa\over2}\langle\bar\varphi\gamma_5\phi-\bar\phi\gamma_5\varphi\rangle$, and the Polyakov
line $\langle L\rangle={1\over N_c}\mbox{tr}\langle\prod_{t=1}^{L_t}U_{0;\vec x,t}\rangle$.
Once again, the results for $N_f=4$ are qualitatively very similar to those
found for $N_f=2$.
Note that $\langle qq\rangle$ is
strictly only an order parameter in the limit $j\to0$; we choose to plot the
ratio $\langle qq\rangle/\mu^2$ since for a degenerate system in which
superfluidity arises as a result of BCS condensation
the order parameter is expected to scale
as the area of the Fermi surface via
$\langle qq\rangle\propto\Delta_{BCS}\mu^2$, where $\Delta_{BCS}$ is the 
superfluid gap.
Fig.~\ref{fig:polyandqq} confirms that this is 
indeed the case for $\mu\in(\mu_Q,\mu_D)$; the data for $N_f=4$ being 
flatter than those for $N_f=2$, although both manifest a shallow minimum at
$\mu/m_\pi\approx0.95$.  This suggests the inequality
\begin{equation}
\Delta_{BCS}^{N_f=4}a<\Delta_{BCS}^{N_f=2}a.
\end{equation}
Whilst this may signal a physical difference between the two theories, the
influence of lattice artifacts, or indeed the opening up of other more favoured condensation
channels in the flavor-rich case, cannot yet be eliminated.

Fig.~\ref{fig:polyandqq} also shows a sharp transition between $\langle L\rangle\approx0$ and
$\langle L\rangle>0$ at $\mu_D\approx1.0m_\pi$. This transition 
approximately coincides
with the upturn in $n_q/n_{SB}^{\rm latt}$ observed in Fig.~\ref{fig:4fln}. Since $\langle
L\rangle\sim\exp(-f_q/T)$, where $f_q$ is the free energy of a
static isolated color source, we interpret $\mu_D$ as the chemical potential at
which color deconfinement takes place. It is striking that $\mu_D^{N_f=2}/m_\pi$ and
$\mu_D^{N_f=4}/m_\pi$ appear to be identical; we attribute the difference in magnitude of
$\langle L\rangle$ for $\mu>\mu_D$ to the smoother gauge
fields for $N_f=4$, resulting in a smaller downwards
renormalisation of the Polyakov line~\cite{Gupta:2007ax}.

We therefore recover for QC$_2$D with $N_f=4$ the same intriguing result found for
$N_f=2$~\cite{Hands:2010gd}; namely that for low temperatures $T\ll\mu$,
$\mu_D>\mu_Q$, implying that there is a phase with the thermodynamic properties
of degenerate quark matter, but in which color is confined. The current result
is if anything stronger than that of \cite{Hands:2010gd} because $n_q(\mu)$
approaches the free quark result much more closely as a result of the smoother
gauge fields. Such a phase is
reminiscent of the confined, chirally-symmetric {\em quarkyonic\/} phase
originally discussed in the context of large-$N_c$ gauge
theories~\cite{McLerran:2007qj}. Unfortunately our use of Wilson lattice
fermions, which have no chiral symmetry away from the limit $\kappa\to\kappa_c$,
precludes a discussion of whether chiral symmetry is restored for $\mu>\mu_Q$ at
present.

\begin{figure}[htb]
\includegraphics*[width=\colw]{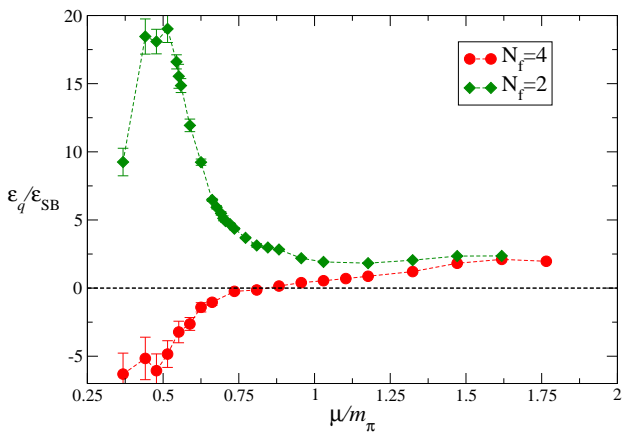}
\caption{\em  $\varepsilon_q/\varepsilon_{SB}^{\rm latt}$ vs. $\mu/m_\pi$
for $N_f=2,4$.}
\label{fig:eps_q}
\end{figure}
So far we have found a close similarity between $N_f=2$ and $N_f=4$. This does not
carry over to the quark energy density $\varepsilon_q$, defined here by
\begin{eqnarray}
\varepsilon_q=\kappa\sum_{i=1}^{N_f}\Bigl\langle
\bar\psi_x^i(\gamma_0&-&1)e^\mu U_{0x}\psi_{x+\hat0}^i-\nonumber\\
\bar\psi_x^i(\gamma_0&+&1)e^{-\mu}
U_{0x-\hat0}^\dagger\psi_{x-\hat0}^i\Bigr\rangle.
\end{eqnarray}
Fig.~\ref{fig:eps_q} plots $\varepsilon_q/\varepsilon_{SB}^{\rm latt}$ versus
$\mu$ for $N_f=2,4$. Note that a vacuum contribution $\varepsilon_q^0$ evaluated
at $\mu=0$ must
be subtracted from both interacting and free data; for $N_f=4$ this
correction $\varepsilon_q^0a^4=0.3724(10)$. 
Even after this additive correction there is still a
multiplicative renormalisation required by a $\mu$-independent factor known as a
Karsch coefficient~\cite{Karsch:1989fu}. Non-perturbative values for Karsch
coefficients are still to be determined for QC$_2$D, but the shapes of the
curves are in principle correct up to discretisation errors.
Fig.~\ref{fig:eps_q} shows a big difference at low values of $\mu$ between
$N_f=2$, where $\varepsilon_q/\varepsilon_{SB}^{\rm latt}$ has a peak considerably larger 
than that predicted by $\chi$PT~\cite{Hands:2006ve,Hands:2010gd}, and $N_f=4$,
where the ratio actually starts negative and rises monotonically with $\mu$.
A negative value of $\varepsilon_q$ is not forbidden {\it a priori}, but
the requirement for positivity of the total energy density certainly constrains the 
contribution $\varepsilon_g$ from the gluons (see below).
For $\mu\gapprox\mu_D$ $\varepsilon_q^{N_f=2}/\varepsilon_{SB}^{\rm
latt}\approx2$ becomes 
approximately  constant; $\varepsilon_q^{N_f=4}/\varepsilon_{SB}^{\rm 
latt}$ approaches this value from below, and for $\mu/m_\pi \gapprox1.5$ the two models
appear to coincide up to the unknown Karsch correction.

\begin{figure}[htb]
\includegraphics*[width=\colw]{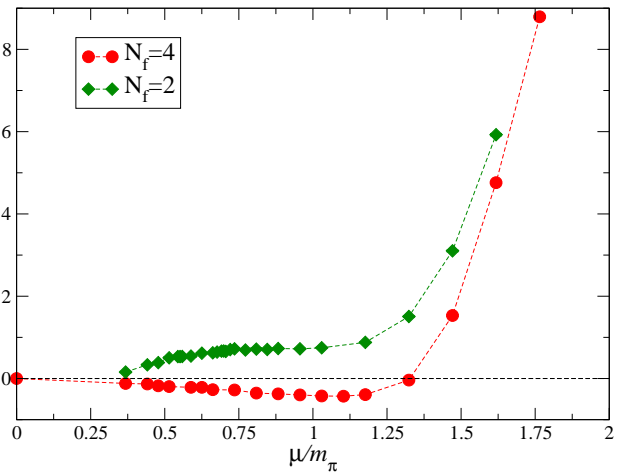}
\caption{\em  The unrenormalised quark contribution to the trace anomaly
$\kappa^{-1}\mbox{Tr}(\One-\langle\bar\psi\psi\rangle)$
vs. $\mu/m_\pi$ for $N_f=2,4$.}
\label{fig:tmumuq}
\end{figure}
A related quantity is the quark contribution to the trace of the stress-energy
tensor $(T_{\mu\mu})_q$, given by
\begin{equation}
(T_{\mu\mu})_q=a{{\partial\kappa}\over{\partial a}}
\times{1\over\kappa}(4N_fN_c-\langle\bar\psi\psi\rangle).
\end{equation}
With data from only one lattice spacing, we are currently unable to estimate the
beta-function;
Fig.~\ref{fig:tmumuq} plots raw values of $\kappa^{-1}\mbox{Tr}(\One-\langle\bar\psi\psi\rangle)$
for $N_f=2,4$, normalised to two quark flavors for ease of comparison, and
including the necessary vacuum subtraction.
Qualitatively they have very different behaviour for $\mu<\mu_D$, and suggest
that $(T_{\mu\mu})_q^{N_f=2}$ and $(T_{\mu\mu})_q^{N_f=4}$ differ even in sign
in this regime. Note that $\chi$PT predicts $T_{\mu\mu}>0$ for
$\mu_o<\mu<\surd3\mu_o$~\cite{Hands:2006ve}.
Since $T_{\mu\mu}=\varepsilon-3p$ for isotropic matter, the
negative sign of $(T_{\mu\mu})_q^{N_f=4}$ is consistent with the negative value of
$\varepsilon_q$ for small $\mu$ reported in the previous paragraph.  
Once $\mu\gapprox\mu_D$, both models exhibit a strong upward trend, suggesting that
quarks dominate $T_{\mu\mu}$ in the deconfined phase.

\begin{figure}[htb]
\includegraphics*[width=\colw]{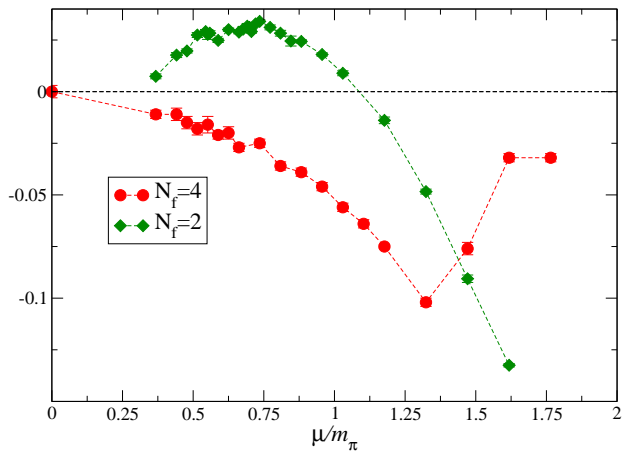}
\caption{\em  The unrenormalised gluon contribution to the trace anomaly
$3\beta(\Box_s+\Box_t)$ vs. $\mu/m_\pi$ for $N_f=2,4$.}
\label{fig:tmumug}
\end{figure}

Finally we present results for local gluonic observables.
In a
non-Lorentz invariant system such as one with $\mu\not=0$ it is helpful to
define
\begin{equation}
\Box_s={1\over3N_c}
\sum_{i<j}\langle\mbox{tr}U_{ijx}\rangle;\;\;\
\Box_t={1\over3N_c}
\sum_x\sum_{i}\langle\mbox{tr}U_{0ix}\rangle.
\end{equation}
\begin{figure}[htb]
\includegraphics*[width=\colw]{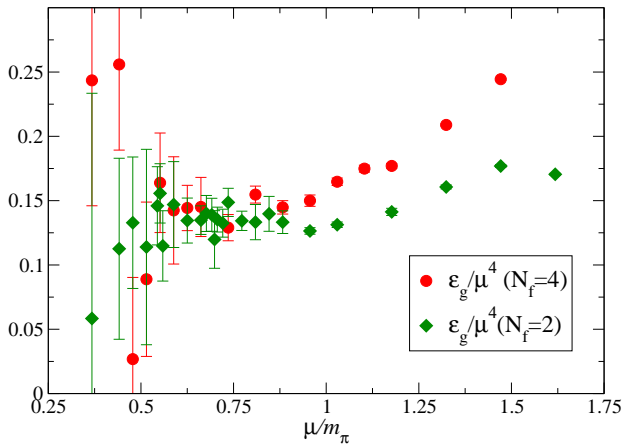}
\caption{\em  $\varepsilon_g/\mu^4$ vs. $\mu/m_\pi$ for $N_f=2,4$.}
\label{fig:4flegbymu4}
\end{figure}
We then consider in Fig.~\ref{fig:4flegbymu4} 
the difference, proportional to the gluon energy density
\begin{equation}
\varepsilon_g=3Z\beta(\Box_t-\Box_s),
\end{equation}
where $Z$ is another as yet undetermined Karsch coefficient (assumed unity in
the figure),
and in Fig.~\ref{fig:tmumug} the average plaquette
related to the gluon component of the stress-energy
tensor via
\begin{equation}
(T_{\mu\mu})_g=-a{{\partial\beta}\over{\partial a}}\times3\beta(\Box_s+\Box_t).
\end{equation}

For the gluon energy density (Fig.~\ref{fig:4flegbymu4})
we have no analytic expectation from either free field theory or $\chi$PT. 
As in previous work, we plot the dimensionless quantity $\varepsilon_g/\mu^4$;
modulo the Karsch coefficient its value for $\mu\lapprox\mu_Q$ is approximately
constant and similar in magnitude
for $N_f=2,4$ (at the smallest $\mu$ the errors are hard to control), but for
$\mu\gapprox\mu_D$ the ratio increases faster for $N_f=4$, and has essentially doubled
by $\mu/m_\pi=1.5$. For $(T_{\mu\mu})_g$ (Fig.~\ref{fig:tmumug}) we note that again a
vacuum subtraction must be applied. 
It is clear that there is a regime
$\mu_o\lapprox\mu\lapprox\mu_D$ where $(T_{\mu\mu})_g^{N_f=2}$ is positive,
before turning over to become negative at large $\mu$; by contrast 
$(T_{\mu\mu})_g^{N_f=4}<0$ throughout. It is not yet clear whether the sharp
upward kink at $\mu/m_\pi =1.3$ is physical, or merely a discretisation artifact.

The data of Figs.~\ref{fig:tmumuq},\ref{fig:tmumug} taken together imply a major
qualitative difference between $N_f=2$ and $N_f=4$: in the former case,
$T_{\mu\mu}>0$ for all $\mu$, which in particular is consistent with the existence
of a non-relativistic description with $\varepsilon\gg p$ for
$\mu\gapprox\mu_o$. For $N_f=4$ by contrast, $T_{\mu\mu}<0$ for $\mu<\mu_D$ (we
are unable to make a firm statement at this point about the deconfined phase due to the
absence of the beta-functions), implying that the matter which forms at onset is
already strongly-interacting and relativistic.

\section{Discussion}
\label{sec:disc}

In this paper we have presented the first exploratory results for
flavor-rich baryonic matter with $N_f=4$ in QC$_2$D. Since available resources
have restricted us to a single lattice spacing, hopping parameter and diquark
source strength, we have chosen to match the simulation to a previous study of
$N_f=2$ on the same lattice size with the same value of $m_\pi
a$~\cite{Hands:2010gd}, implying the
same onset chemical potential $\mu_oa$ measured in lattice units. Accordingly
the comparison between the two theories is best performed using observables
measured in units of $m_\pi$. Since $\sigma^{N_f=4}a^2<\sigma^{N_f=2}a^2$,
we expect the lattice gauge fields to be much smoother in the flavor-rich case,
although that renders results more susceptible to finite-volume and thermal
corrections due to the corresponding reduction in $L_s\surd\sigma$,
$L_t\surd\sigma$. One particular concern in the latter case is that the static
potential $V(R)$ used to calibrate the lattice 
could actually be underestimated due to
thermal screening. 
Another caveat 
which must eventually be addressed is control over the extrapolation to 
continuum $a\to0$ and zero source $j\to0$ limits, currently being addressed for
$N_f=2$~\cite{SJHJIS}. 

Let us cite one example of a potential issue in the current study, the total energy density
$\varepsilon$. If we make the most optimistic assumption
that the Karsch coefficients are approximately unity, and that both $j\to0$ and
$a\to0$ extrapolations will not change the results much, then the data of
Figs.~\ref{fig:eps_q},\ref{fig:4flegbymu4} suggest that for $\mu\sim\mu_o$ with $N_f=4$
$\varepsilon/\mu^4=\varepsilon_q/\mu^4+\varepsilon_g/\mu^4\approx-0.7+0.13<0$,
which is clearly unphysical.

We also note that at $\mu/m_\pi=1.76$, the largest chemical
potential studied, the ratio $n_q^{N_f=4}/n^{\rm sat}=0.32$, where $n^{\rm sat}=2N_fN_c$
is the saturation density obtained when every lattice site is maximally occupied 
by fermionic degrees of freedom; hence we might be concerned that in this
high-density regime the results are susceptible to saturation artifacts.
In this light, the superfluid condensate in Fig.~\ref{fig:polyandqq} and gluon
contribution to the stress-energy tensor in Fig.~\ref{fig:tmumug} are especially
troubling, since both show a sharp kink at $\mu/m_\pi=1.3$, for which we currently have
no explanation.

Bearing these caveats in mind, let us review the principal results.
Fig.~\ref{fig:4fln} for $n_q$, $p$ for $N_f=2,4$ 
expressed as fractions of the free-field
results, demonstrates that both 
models display a qualitatively similar sequence of transitions as $\mu$ is
increased. For $N_f=4$ there is a range $0.75<\mu/m_\pi<1.0$ where
$n_q/n_{SB}^{\rm latt}\approx1$, implying the existence of a
well-defined Fermi surface. Further study will be needed to establish whether
the difference in the height of this plateau results from the diminished
impact of lattice artifacts in the $N_f=4$ simulation, or a genuine physical
difference between the two theories. Still, the enhanced flatness of the $N_f=4$
plateau and the plausibility of the resulting physical picture
increases our confidence in the slightly {\it ad hoc\/} procedure we
have used to compensate for discretisation artifacts.

Another result is the confirmation of the major finding of ~\cite{Hands:2010gd},
namely that $\mu_D>\mu_Q$, or in other words, for both $N_f=2$ and 4
there is a range of $\mu$ where
degenerate quark matter remains color-confined.
It will be very interesting to elucidate the nature of the deconfined phase
further, since at first sight it bears little resemblance to thermal
deconfinement. For instance, Fig.~\ref{fig:4flegbymu4} suggests that the gluon
energy density rises smoothly for $\mu\gapprox\mu_D$, with $\varepsilon_g/\mu^4$
increasing by at most a factor of two over its value near onset.

Finally, the most interesting and unexpected feature of the new simulation is the big
difference between $N_f=2,4$ for the quark contributions to the
energy density (Fig.~\ref{fig:eps_q}) 
and stress-energy tensor
(Fig.~\ref{fig:tmumuq}). For moderate $\mu$ both $\varepsilon_q^{N_f=4}$ and
$(T_{\mu\mu})_q^{N_f=4}$ are negative; this result is self consistent and
independent of the value of the relevant Karsch coefficient, though as discussed
above could potentially alter as the limits $a\to0$, $j\to0$ are taken.
The disparity between $N_f=2$ and 4 however, and of each with the predictions of
$\chi$PT~\cite{Hands:2006ve}, is striking. It suggests very different physical
descriptions in the low-density regime: as $\mu\to\mu_{o+}$ the $N_f=2$
theory appears to be a non-relativistic BEC formed of weakly-interacting 
tightly-bound diquark
bosons, consistent with $\chi$PT~\cite{Kogut:2000ek} and
yielding $T_{\mu\mu}>0$, whereas with $N_f=4$ the matter appears to
be relativistic
and strongly-interacting for all $\mu>\mu_o$.

\begin{acknowledgement}
This project was enabled with the assistance of IBM Deep Computing.
S.K. was supported by the National Research Foundation of Korea grant funded 
by the Korea government (MEST) No. 2010-0022219.
\end{acknowledgement}

\end{document}